\documentstyle[aps,prl,epsfig]{revtex}

\def\mps{M_{\rm PS}(t)}

\def\mintt{M_{\rm int}(t)}

\def\mextt{M_{\rm ext}(t)}
\def\mmax{M_{\rm int}^{\rm max}}
\def\ton{t_{\rm on}}
\def\toff{t_{\rm of\/f}}
\def\locl{l_{\infty}}

\begin{document}
\draft

\twocolumn[\hsize\textwidth\columnwidth\hsize\csname @twocolumnfalse\endcsname

\title{Superballistic spreading of wave packets}
\author{L. Hufnagel, R. Ketzmerick, T. Kottos, and T. Geisel}
\address{ Max-Planck-Institut f\"ur Str\"omungsforschung
und Institut f\"ur Nichtlineare Dynamik der Universit\"at G\"ottingen,\\
Bunsenstra{\ss}e~10, 37073 G\"ottingen, Germany}

\date{\today}

\maketitle

\begin{abstract}
We demonstrate for various systems that the variance of a wave packet 
$M(t)\propto t^\nu$, can show a {\it superballistic} increase with $2<\nu\le3$,
for parametrically large time intervals. A model is constructed which explains 
this phenomenon and its predictions are verified numerically for various 
disordered and quasi-periodic systems.
\end{abstract}


]

\begin{figure}
\begin{center}
    \epsfxsize=8.5cm
    \leavevmode
    \epsffile{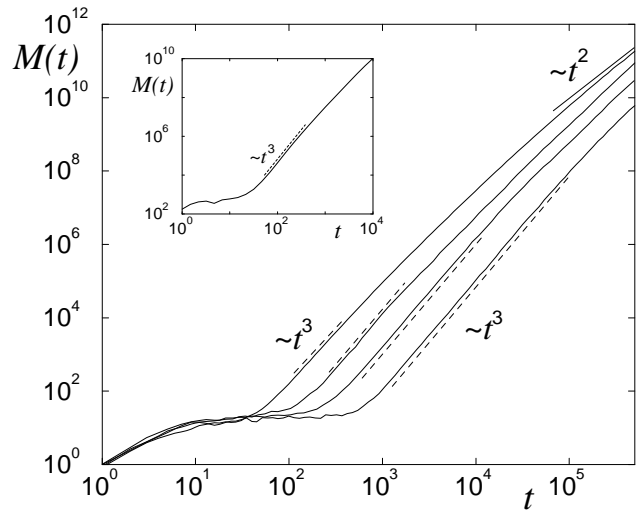}
\caption[model]{
Variance $M(t)$ of a wave packet in a 1D disordered model of size $L=50$, 70, 80, 100
(from left to right) with a perfect lattice attached to both ends.
The dashed lines indicate the time range over which the cubic growth appears
before the asymptotic ballistic spreading sets in.
The inset shows the cubic growth for a Band Random Matrix model ($b=10$, $L=100$, $T=0.001$).
}
\label{fig1}
\end{center}
\end{figure}

The time evolution of wave packets in one dimensional (1D) and quasi-1D
lattices is described by the time-dependent Schr\"odinger equation
\begin{equation}
\label{schr}
i\,{\frac{dc_n(t)}{{dt}}}=\,\sum_{m=n-b}^{n+b}H_{nm}c_m\quad ,
\end{equation}
where $c_n(t)$ is the probability amplitude for an electron to be at site $n$,
$b$ is the number of channels, and $H_{nm}$ is a tight-binding Hamiltonian.
When translational symmetry is present, the eigenstates of the system are 
plane waves and the variance of wave packets increases quadratically in time
("ballistic spreading"). On the other hand, since the pioneering work of 
Anderson \cite{A58}, it is known that disorder usually tends to suppress 
propagation and leads to localization. In the one-dimensional case, even a 
small amount of disorder leads to localization of all eigenstates\cite{A58,book}, 
and therefore asymptotically the spreading of a wave packet remains bounded. 
In higher dimensions a localization to delocalization transition can occur
which leads to diffusive or subdiffusive spreading of a wave packet\cite{book,HS94}.
In addition, there are quasi-periodic systems which even in 1D show fractal 
energy spectra and eigenfunctions leading to a power law spreading \cite{KKKG97} 
that is reminiscent of anomalous diffusion in classical systems.
A global characterization of the dynamical evolution of a wave packet is provided 
by its variance
\begin{equation}\label{defvar}
M(t)\equiv\sum_nn^2|c_n(t)|^2 \propto t^{\nu} \quad .
\end{equation}
Its time dependence gives a quantitative description of the dynamics:
$\nu=0$ corresponds to localization, $\nu=1$ to diffusion, $\nu=2$ to
ballistic motion, and $\nu\in(0,2)$ to anomalous diffusion.
It was shown that a ballistic upper bound, $M(t)\le A\cdot t^2$, exists for all
times with a system specific constant $A$\cite{statement}.
Although this statement gives no restriction on $\nu$ for
finite time intervals, to our knowledge in all studies up to now $\nu\le2$
was found for any time.
In this paper we show that there can be superballistic spreading with exponents 
$\nu\in(2,3]$
for parametrically large time intervals.
Examples with $\nu=3$ can be seen in Fig.~\ref{fig1} where the variances of 
wave packets, while staying below the ballistic upper bound, show a cubic growth.
Each system consists of a perfect lattice with a disordered region of finite length.
Moreover, extending the length of the disordered region {\it increases} the time interval 
of the cubic growth.
We will show that this unexpected behavior of the variance is associated with the rate
$\Gamma$ for the emission of an electron from the disordered region into the perfect lattice.
This parameter determines the time scales
\begin{equation}
\label{tonoff}
\ton\sim \left({1\over \Gamma}\right)^{1/\nu}\quad{\rm and}\quad\toff\sim{1\over\Gamma}\quad,
\end{equation}
when the superballistic growth starts and ends, respectively. The different 
$\Gamma-$dependence of $\ton$ and $\toff$ ensures that this time interval becomes 
arbitrarily large as $\Gamma$ decreases. We would like to mention that such
intermediate superballistic regimes are well known in other contexts like in 
hydrodynamic turbulence (Richardson law) or in plasma physics, but are unrelated.
In order to understand the appearance of the cubic growth of the variance in
Fig.~\ref{fig1} we consider a simple probabilistic model:
The disordered part is replaced by a point source and anything emitted from it
moves with a constant velocity $v$ modelling the dynamics of a perfect lattice.
Initially, all probability is confined to the point source and decays with a constant
rate $\Gamma$, such that the probability at the point source is given by
\begin{equation}\label{eqstay}
P(t) = \exp{(-\Gamma t)}.
\end{equation}
The variance $\mps$ of the point-source model is then given by
\begin{equation}\label{eqmps}
\mps = \int_0^\infty\!\!\!\!\!dx x^2\int_0^t\!\!\!dt^{'}
(-\dot{P}(t^{'}))\:\delta(x-v(t-t^{'}))\quad,
\end{equation}
where $-\dot{P}(t)$ is the flux emitted from the point source.
Substituting Eq.~(\ref{eqstay}) for $P(t)$ we get
\begin{equation}\label{eqmpspoft}
\mps = v^2\Gamma\int_0^t dt^{'} e^{-\Gamma t^{'}} (t-t^{'})^2\quad,
\end{equation}
which yields after integration
\begin{equation}
\label{var3}
\mps= v^2\left( t^2- {2\over \Gamma} t + {2\over \Gamma^2}
-{2\over \Gamma^2} e^{-\Gamma t}\right)\quad.
\end{equation}
As expected, the variance grows quadratically, $\mps\sim v^2 t^2$,
for asymptotically large times.
Expanding the exponential term in Eq.~(\ref{var3}) one finds
\begin{equation}
\label{varfin}
\mps = v^2\Gamma\left({1 \over 3} t^3-{\Gamma\over 12} t^4+...\right)\quad.
\end{equation}
Under the condition $t<1/\Gamma$ the cubic term dominates all higher orders.
Thus we find a cubic increase of the variance in the point-source model
starting from the time $\ton=0$ up to the time $\toff\approx 1/\Gamma$.
At the same time scale the crossover to the asymptotic ballistic spreading
starts, as can be seen from Eq.~(\ref{var3}).
An intuitive understanding is based on the fact, that the linear decrease of $P(t)=1-\Gamma t$
for small times is responsible for the cubic growth of the variance [Eq.~(\ref{eqmps})].
During this time the norm of the wave packet outside the point source increases linearly. This 
linear increase of the norm combined with the usual quadratic increase of the variance 
due to the ballistic spreading yields the cubic growth of the variance.
A more realistic model should take into account the length $L$ of the disordered region
and the time $\tau$ at which the norm of the wave packet outside the disordered region 
starts to increase linearly.
The total variance is $M(t) = \mintt + \mextt$, where $\mintt$  is the contribution
of the internal region and $\mextt$ originates from the perfect lattice, which is given by
\begin{equation}\label{eqmextps}
\mextt=\int_L^\infty\!\!\!\!\!dx x^2\int_\tau^t\!\!\!\!dt^{'}
(-\dot{P}(t^{'}))\:\delta(x\!-\!L-\!v(t-t^{'}\!-\!\tau)).
\end{equation}
This reduces to Eq.~(\ref{eqmps}) of the point-source model
for times $t$ fulfilling the following three conditions:
(i) $t\gg\tau$, (ii) $vt\gg L$, and (iii) $\mextt\gg\mintt$.
These conditions set the time scale $\ton$ at which the cubic law may start, in contrast
to $\ton=0$ for the point-source model.
Conditions (i) and (ii) give the time scales $\tau$ and $L/v$, respectively.
The internal variance $\mintt$ is bounded by $\mmax\le L^2$.
Thus condition (iii) together with $\mextt\approx v^2\Gamma/3t^3$
valid for $t<\toff$ leads to the time scale $(\mmax/(v^2\Gamma))^{1/3}$.
The maximum of these three time scales defines the onset of the cubic law
\begin{equation}\label{eqton}
\ton= {\rm max}\{\tau,\frac{L}{v},\left(\frac{\mmax}{v^2\Gamma}\right)^{1/3}\}\quad.
\end{equation}
In order to see the cubic law over a large time interval $\toff\approx 1/\Gamma$ 
has to be large implying a small $\Gamma$.
For sufficiently small $\Gamma$ the third time scale in Eq.~(\ref{eqton})
will dominate leading to the scaling presented in Eq.~(\ref{tonoff}) for $\nu=3$.
The ratio $\toff/\ton$ scales as $\Gamma^{-2/3}$ and therefore can be made 
arbitrarily large by decreasing $\Gamma$.
\begin{figure}
\begin{center}
\hspace*{-0.5cm}\epsfig{figure=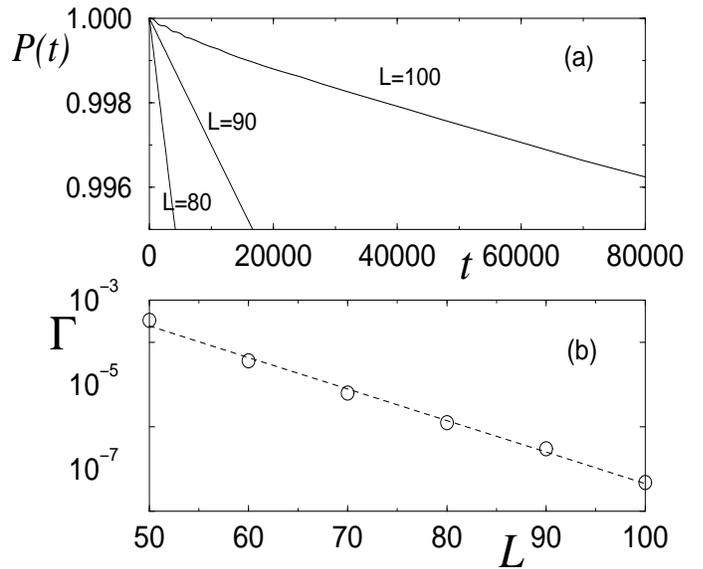,height=9cm,width=8cm,angle=270}
\caption[model]{
(a) Norm of the wave packet inside the disordered region of the 1D model vs. t
for $L=80$, 90, 100 showing linear decays.
(b) Decay rate $\Gamma$ vs. $L$ following an exponential law (dashed line).
}
\label{fig2}
\end{center}
\end{figure}
One can extend the above analysis to other dynamical exponents $\nu<3$
by embedding the internal region in a lattice showing anomalous diffusion. 
These lattices are characterized by a variance scaling as $t^{\mu}$, $\mu\in(0,2)$.
This yields the general expression $\nu=\mu+1$ for Eqs.~(\ref{defvar},\ref{tonoff}).
Again, the intuitive understanding is that the linear increase of the norm of the 
wave packet outside the internal region increases the exponent by 1.
\begin{figure}
\begin{center}
    \epsfxsize=8.5cm
    \leavevmode
    \epsffile{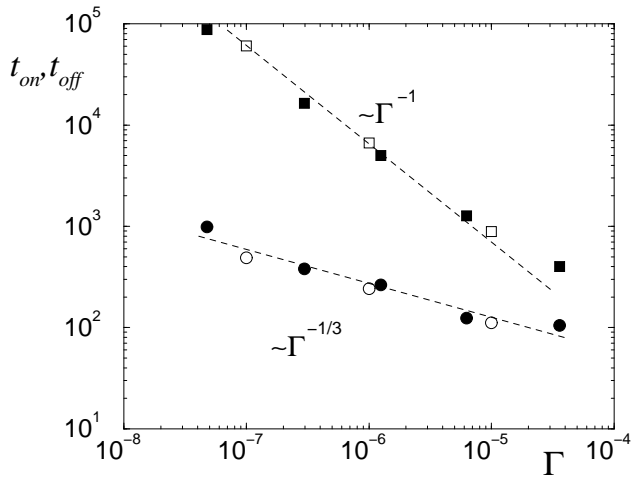}
\caption[model]{
Times $\ton$ (circles) and $\toff$ (squares) vs. $\Gamma$
for the 1D model (filled symbols) and for the Band Random Matrix
model (open symbols, data for $\ton$ and $\toff$ are shifted to account
for different prefactors of the power laws for the two models).
}
\label{fig3}
\end{center}
\end{figure}
In the remainder of this paper we will give numerical evidence supporting the
above analysis.
We use perfect or quasi-periodic lattices with a finite 1D or quasi-1D
disordered region, although there are many other possible settings.
Our first example consists of a 1D disordered
region of size $L$ attached with semi-infinite perfect lattices at both ends.
In Eq.~(\ref{schr}) this corresponds to a tridiagonal Hamiltonian $(b=1)$,
with $H_{n,n\pm 1}=1$ and $H_{nn}=0$ except for a region of size $L$ where
$H_{nn}$ is a random number. We fix the disorder strength and we use sample 
sizes $L=50,\dots,100$ such that $L>\locl$, where $\locl$ is the localization
length of the corresponding infinite disordered system.
In all cases the initial $\delta-$like wave packet is launched in the middle of the
disordered region.
Fig.~\ref{fig1} shows the variance averaged over 10 disorder realizations.
It should be noted, that without averaging we get the same qualitative behavior.
For small times all wave packets spread ballistically until the variance starts to
saturate when $\mintt\approx\locl^2$.
Then the cubic increase of the variance can be observed before finally the asymptotic
ballistic spreading sets in.
There is a ballistic upper bound for all times in agreement with Ref.~\cite{statement}.
The range over which the cubic law holds increases with the size of the disordered sample.
This is in agreement with the predictions of the point-source model:
It is applicable since the norm of the wave packet inside the disordered region initially decays
linearly (Fig.~\ref{fig2}a).
From this linear decay $P(t)=1-\Gamma t$ we determine the rate $\Gamma$ which decreases
exponentially with the sample size $L$ (Fig.~\ref{fig2}b), as expected for the localized regime\cite{BGZ98}.
Thus extending the disordered region decreases $\Gamma$ and together with Eq.~(\ref{tonoff}) explains
the increase of the superballistic time interval.
The time scales $\ton$ and $\toff $ are determined from the times when $M(t)$ deviates
from the fitted cubic increase by more than 10\%.
The scaling of the time scales $\ton$ and $\toff$ with $\Gamma$ are shown in Fig.~\ref{fig3} 
confirming Eq.~(\ref{tonoff}).
We now show that our considerations also apply to a Band Random Matrix model that
describes quantum wires\cite{FM94}.
The Hamiltonian matrix $H_{nm}$ is real and the entries are different from
zero in a stripe of width $b$ around the diagonal ($|n-m|\le b$), only.
The parameter~$b$ defines the hopping range between neighbouring sites, or,
in the quasi-1D interpretation, the number of transverse channels along a thin
wire\cite{FM94}.
The non-zero matrix elements are independent Gaussian random numbers with variance
1 within a region of size $L$ and are equal to 1 outside this region.
The matrix elements $H_{nm}$ that couple the sample to the perfect lattice
are random numbers with variance $T$ such that $\Gamma\sim T$\cite{weiden93}.
In order to study an example where the internal variance $\mintt$ is bounded by the
sample size rather than the localization length, we choose a sample size $L\le\locl$.
Thus the initial wave packet spreads diffusively over the disordered region before it
leaks out to the leads.
We find a cubic increase of the variance (Fig.~\ref{fig1}, inset) and by varying the
coupling strength $T$ we confirm Eq.~(\ref{tonoff}) for $\ton$ and $\toff$ (Fig.~\ref{fig3}).
Finally, we use a Fibonacci chain model\cite{firstfibo} outside the disordered region,
which allows to vary the exponent $\mu$ of its anomalous diffusion by changing the
potential strength $V$\cite{GKP95}.
Fig.~\ref{fig4} shows the superballistic increase of the external variance in nice agreement
with the expected exponent $\nu=\mu+1$.
\begin{figure}
\begin{center}
    \epsfxsize=8.5cm
    \leavevmode
    \epsffile{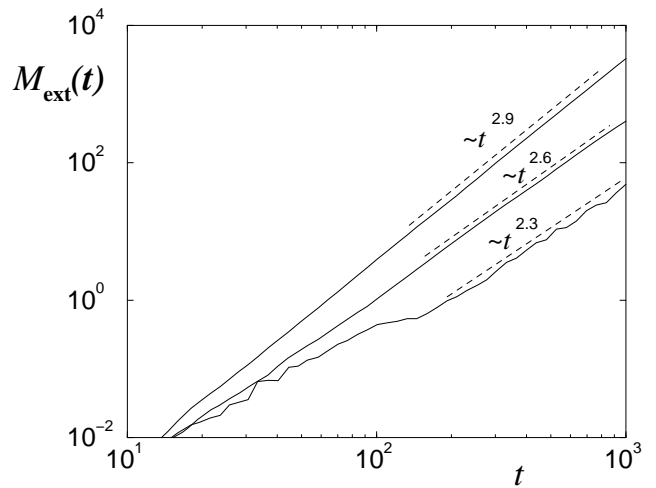}
\caption[model]{
External variance $\mextt$ for a 1D disordered sample ($L=50$) attached to Fibonacci 
chain models with $V=0.1$, 0.4, and 0.7 (from top to bottom), showing the expected 
power laws (dashed lines).
}
\label{fig4}
\end{center}
\end{figure}
We thank R.~Fleischmann, I.~Guarneri, and A.~Politi for helpful discussions.


\begin{thebibliography}{99}

%
\bibitem{A58} P.~W.~Anderson, Phys. Rev. {\bf 109}, 1492 (1958).
%
\bibitem{book} E.~N.~ Economou, {\it Green's Functions in Quantum
Physics}, Springer Series in Solid State Physics, Vol. 7
Springer-Verlag, Berlin, (1979).
B. Kramer and A. Mackinnon, Rep. Prog. Phys. {\bf 56},1469 (1993).
%
\bibitem{HS94} B. Huckestein and L. Schweitzer, Phys. Rev. Lett. {\bf 72},
713 (1994);
%
\bibitem{KKKG97} I.~Guarneri, Europhys. Lett. {\bf 10}, 95 (1989);
ibid {\bf 21}, 729 (1993);
R. Ketzmerick, K. Kruse, S. Kraut, and T. Geisel, Phys. Rev. Lett. {\bf 79},
1959 (1997).
%
\bibitem{statement} I.~Guarneri, J. Math. Phys., {\bf 37}, 5195 (1996);
I.~Guarneri and H.~Schulz-Baldes, Rev. Math. Phys. {\bf 11}, 1249, (1999).
%
\bibitem{BGZ98} A. Buchleitner, I. Guarneri, and J. Zakrzewski, Europhys. Lett. 
{\bf 44}, 162 (1998);
O. A. Starykh, P. R. J. Jacquod, E.~E.~Narimanov, and A. D. Stone, cond-mat/0001017 (2000).
%
\bibitem{FM94} Y. F. Fyodorov and A. D. Mirlin, Int.J.Mod.Phys., {\bf 8} 3795 (1994);
T. Kottos, A.~Politi, and G.~P.~Tsironis, Phys.~Rev.~E, {\bf 55}, 4951 (1997);
M. Weiss, T. Kottos, and T. Geisel, Phys. Rev. B, in press; [cond-mat/0005339].
%
\bibitem{weiden93} H.~A.~Weidenm\"uller, Physica A, {\bf 167}, 28, (1990).
%
\bibitem{firstfibo} M.~Kohmoto, L.~P.~Kadanoff, and C.~Tang, Phys. Rev. Lett. {\bf 50}, 1870 (1983);
S.~Ostlund, R.~Pandit, D.~Rand, H.~J.~Schellnhuber, and E.~D.~Siggia, Phys. Rev. Lett. {\bf 50}, 1873 (1983).
%
\bibitem{GKP95}T. Geisel, R. Ketzmerick, and G. Petschel, in {\it Quantum Chaos;
Between Order and Disorder}, G. Casati and B. Chirikov, eds. (Cambridge University Press, Cambridge, 1995),
page 634, (1995).
%
\end{thebibliography}
\end{document}